\documentclass{sig-alternate-2013}

\permission{\copyright 2017 International World Wide Web Conference Committee \\ (IW3C2), published under Creative Commons CC BY 4.0 License.}
\conferenceinfo{WWW 2017,}{April 3--7, 2017, Perth, Australia.}
\copyrightetc{ACM \the\acmcopyr}
\crdata{978-1-4503-4913-0/17/04. \\
http://dx.doi.org/10.1145/3038912.3052637  \\
\includegraphics{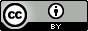}
}

\usepackage{color}
\usepackage{url}
\usepackage{verbatim}
\usepackage{subfigure}

\usepackage{multirow}
\usepackage{algorithm}
\usepackage[noend]{algorithmic}
\usepackage{xspace}
\usepackage{graphicx}
\usepackage{epsfig}
\usepackage{amsmath}
\usepackage{amssymb}
\usepackage{times}
\usepackage{float}
\usepackage{enumitem}
\usepackage{balance}
\usepackage[font=bf,belowskip=0pt,aboveskip=5pt]{caption}
\usepackage{lipsum}

\usepackage{amsmath}
\makeatletter
\g@addto@macro\normalsize{%
  \setlength\abovedisplayskip{2pt}
  \setlength\belowdisplayskip{2pt}
  \setlength\abovedisplayshortskip{2pt}
  \setlength\belowdisplayshortskip{2pt}
}

\usepackage{epigraph}

\usepackage{booktabs}

\usepackage{paralist}
\renewenvironment{itemize}[1]{\begin{compactitem}#1}{\end{compactitem}}

\newcommand{\hide}[1]{}
\newcommand{\xhdr}[1]{\vspace{1.7mm}\noindent{{\bf #1.}}}

\newcommand{\etc}{\emph{etc.}}
\newcommand{\eg}{\emph{e.g.}}
\newcommand{\ie}{\emph{i.e.}}

\graphicspath{{./FIG/}}

\begin{document}

\title{Harnessing the Web for Population-Scale Physiological Sensing: A Case Study of Sleep and Performance}




\numberofauthors{1}
\author{
\begin{tabular}{ccc}
Tim Althoff\thanks{Research done during an internship at Microsoft Research.} & Eric Horvitz \hspace{0.125cm} Ryen W. White & Jamie Zeitzer\\
\affaddr{Stanford University} & \affaddr{Microsoft Research}  & \begin{tabular}{@{}c@{}} \affaddr{Stanford Center for Sleep} \\ \affaddr{Sciences and Medicine} \end{tabular}\\
\email{althoff@cs.stanford.edu} & \email{\{horvitz,~ryenw\}@microsoft.com} & \email{jzeitzer@stanford.edu}
\end{tabular}
}

\maketitle

\begin{abstract}

Human cognitive performance is critical to productivity, learning, and accident avoidance. Cognitive performance varies throughout each day and is in part driven by intrinsic, near 24-hour circadian rhythms. Prior research on the impact of sleep and circadian rhythms on cognitive performance has typically been restricted to small-scale laboratory-based studies that do not capture the variability of real-world conditions, such as environmental factors, motivation, and sleep patterns in real-world settings. Given these limitations, leading sleep researchers have called for larger \textit{in situ} monitoring of sleep and performance~\cite{roenneberg2013chronobiology}. 
We present the largest study to date on the impact of objectively measured real-world sleep on performance enabled through a reframing of everyday interactions with a web search engine as a series of performance tasks.
Our analysis includes 3 million nights of sleep and 75 million interaction tasks.
We measure cognitive performance through the speed of keystroke and click interactions on a web search engine and correlate them to wearable device-defined sleep measures over time. We demonstrate that real-world performance varies throughout the day and is influenced by both circadian rhythms, chronotype (morning/evening preference), and prior sleep duration and timing. We develop a statistical model that operationalizes a large body of work on sleep and performance and demonstrates that our estimates of circadian rhythms, homeostatic sleep drive, and sleep inertia align with expectations from laboratory-based sleep studies. Further, we quantify the impact of insufficient sleep on real-world performance and show that two consecutive nights with less than six hours of sleep are associated with decreases in performance which last for a period of six days. 
This work demonstrates the feasibility of using online interactions for large-scale physiological sensing.

\end{abstract}


\section{Introduction}

Maintaining optimal cognitive performance has been found to be important in learning~\cite{kelley2015synchronizing}, productivity \cite{colten2006sleep}, and avoiding industrial and motor vehicle accidents~\cite{colten2006sleep,dinges1995overview}. 
Studies have demonstrated that cognitive performance varies throughout the day~\cite{van2000circadian},
likely influencing the quality of our efforts and engagements--in\-clud\-ing how we use and interact with vehicles, devices, resources, and applications. 
Furthermore, cognitive performance is decreased significantly after loss of sleep~\cite{dinges1995overview}. 
Understanding the real-world impact of sleep deficiency is critical. It has been estimated that the cost of fatigue to U.S. businesses exceeds \$150 billion a year in absenteeism, presenteeism, 
workplace accidents, poor and delayed decision-making and other lost productivity on top of the increased health care costs and risk of disease~\cite{hemp2004presenteeism}. 
Despite the important influences, temporal variations of real-world performance are not well understood and have never been characterized on a large scale~\cite{roenneberg2013chronobiology}.

Models of daily patterns in human cognitive performance rely typically on representations of three biological processes: 
\textit{circadian rhythms} (time-dependent, behavior-independent, near 24-hour oscillations)  \cite{van2000circadian}, 
\textit{homeostatic sleep pressure} (the longer awake, the more tired you become)~\cite{borbely1982two}, 
and \textit{sleep inertia} (performance impairment experienced immediately after waking up)~\cite{aakerstedt1997three,dinges1990you}.

While models of these biological processes capture well the patterns of cognitive performance in the  laboratory~\cite{aakerstedt1997three,borbely1982two}, they are based on 
experimental studies in which participants are deprived of sleep and undertake regular, artificial tasks to measure performance instead of non-intrusively capturing performance through everyday tasks in real-world environments. 
In addition, these studies typically include participants that fit a specific physical and psychological profile (\eg, those with depressed mood are often excluded).
Further, participants in an artificial setting can be influenced by their understanding of the study and subconsciously change their behavior to fit the interpretation of its motivation and goals~\cite{orne1962social}.
While laboratory studies have been critical in developing understandings of the basic biological processes that underlie cognitive performance, they fail to account for myriad influences in the real-world, including motivation, mood, illness, environmental conditions, behavioral compensation including caffeine intake, and sleep patterns in the wild that are far more complicated than those enforced in research studies.
How these and other factors alter real-world cognitive performance is not well understood.
Therefore, sleep scientists have called for large-scale real-world measurements of performance and sleep as a necessary step to ``to transform our understanding of sleep'' and ``to establish how to manage sleep to improve productivity, health and quality of life''~\cite{roenneberg2013chronobiology}.

\xhdr{This Work}
We respond to the appeal from the sleep research community with a large-scale study of sleep and performance 
enabled through reframing everyday interactions with a web search engine as a series of performance tasks. 
In particular, we use individual keystrokes when typing a search query and the clicks on search results as a source of precisely timed interactions.
We demonstrate that the timing of these interactions varies based on biological processes and can be used to study the influence of different quantities of sleep on performance.
Search engine interactions offer insight about real-world cognitive performance as they are an integral part of many people's lives and work every day.
More than 90\% of US online adults use web search engines, which now handle billions of searches each day~\cite{Pew2011searchusers}.

Our dataset comprises over 3 million nights of sleep tracked by wearable sensors from 31 thousand users over a period of 18 months and 75 million subsequent real-world performance measurements based on keystrokes and clicks within a web search engine (Section~\ref{sec:dataset}).
This constitutes the largest prospective study of real-world human performance and sleep to date (more than 400 times larger than the second largest comparable study which had only 76 participants~\cite{lim2010meta}).

We first demonstrate that real-world human cognitive performance captured through search engine interactions varies throughout the day in a daily rhythm (Section~\ref{sec:search_performance_measures}).
We find that performance is lowest during habitual sleep times when it is reduced by up to 31\%.
Both the shape and magnitude of this temporal variation are consistent with controlled laboratory-based studies, providing validation of our large-scale performance measures.
We also show that performance varies based on chronotype (morning/evening preference) with early risers performing slowest at 04:00~h (4am) and late risers performing slowest at 07:00~h. 

We then develop a statistical model based on chronobiological research and demonstrate that it successfully disentangles circadian rhythms, homeostatic sleep drive, sleep inertia, and prior sleep duration---key factors considered in the sleep literature (Section~\ref{sec:performance_modeling}).
We quantify that performance varies by 23\% based on time of day, by 19\% based on time since wake up, and by 5\% based on sleep duration (Section~\ref{subsec:model_estimates_results}).
We validate our methodology by demonstrating close agreement between our model estimates based on a large amount of  performance measurements in the wild and smaller controlled sleep studies in artificial laboratory settings. 

After validating our approach, we extend prior laboratory-based sleep research through estimates of how sleep impacts performance in real-world settings. 
In particular, we quantify the impact of one or multiple nights of insufficient sleep on real-world performance (Section~\ref{sec:bad_sleep_impact}).
We demonstrate that very short and very long sleep durations, and irregular timing of sleep are associated with 3\%, 4\% and 7\% lower performance, respectively. 
We also show that two consecutive nights with fewer than six hours of sleep are associated with significantly decreased performance for a period of six days.

Our study is also the first to demonstrate that ambient streams of data, such as patterns of interactions with devices, can be harnessed as large-scale physiological sensors to study and continuously and non-intrusively monitor human performance at population scale. 
The insights and methodology developed in this work are relevant to sleep scientists in pursuit of larger-scale real-world measurements of performance, to computer scientists who build tools and applications that may be affected by variations in human performance, and to the growing community of researchers who have been exploring uses of data from online activities to address questions and challenges in the realm of public health.

\section{Related Work}
\label{sec:related_work}

\xhdr{Circadian Processes in Sleep and Performance}
Empirical studies have found daily rhythms in human performance including alertness, attention, reaction time, memory, and higher executive functions such as planning~\cite{blatter2007circadian}.
The daily variations in performance have been found to be modulated primarily by two processes~\cite{dijk1992circadian}: a \textit{circadian rhythm} (time-dependent, behavior-independent, near 24-hour oscillations)~\cite{van2000circadian} and a \textit{homeostatic sleep drive} (the longer awake, the more tired we become and the more we sleep, the less tired we become)~\cite{borbely1982two}. 
The circadian rhythm acts in opposition to the homeostatic drive for sleep that accumulates across the day, enabling a single, consolidated period of wakefulness throughout the day.
A third process has been proposed called \textit{sleep inertia}~\cite{van2000circadian}, which corresponds to the performance impairment experienced immediately after waking up~\cite{aakerstedt1997three,dinges1990you}. 
In addition to the influence of daily rhythms on the structure of sleep and performance, there are also shorter, 90-minute oscillations, \textit{ultradian rhythms}, that organize the occurrence of NREM and REM stages during sleep.
Ultradian rhythms, circadian rhythms, and homeostatic sleep pressure can all impact the structure, and likely function, of sleep~\cite{dijk1995contribution}.

Human preferences and natural tendency in the relative timing of sleep and wake are called \textit{chronotypes} and are at least partly based on genetics~\cite{roenneberg2003life}. 
Cognitive performance depends on chronotype and time of day~\cite{matchock2009chronotype}; that is, early/morning types (``lark'') tend to be higher performing earlier in the day while late/evening types (``owl'') are higher performing later.
Sleep deprivation has been linked to significant decreases in cognitive performance
that lead to increased risk for accidents and injury~\cite{dinges1995overview}.

A recent study correlated performance on cognitive exercises with a sleep measure based on retrospective self-reports of ``typical sleep'' in 160 thousand users~\cite{sternberg2013largest}.
However, this measure suffers from potential biases~\cite{lauderdale2008self} and does not enable the study of performance variation over time based on time of day and sleep timing. 
Another study showed that insomnia with short sleep is associated with cognitive deficits in 678 subjects~\cite{fernandez2010insomnia} but only measured a single night of sleep to characterize typical sleep patterns after taking performance measurements, leading to similar limitations.
According to a recent meta-analysis~\cite{lim2010meta}, the largest study that measured both sleep
and performance concurrently had 76 participants.

\xhdr{Technology Use and Interaction Patterns} 
Interaction patterns of different devices and applications have been studied on small scale to better understand mobile device usage~\cite{bohmer2011falling}, to detect stress~\cite{vizer2009automated}, used as biometric signals for authentication~\cite{monrose1997authentication}, and linked to biological processes~\cite{murnane2015social,murnane2016mobile} including alertness~\cite{abdullah2016cognitive}.
For example, less sleep was linked to shorter duration of focus of attention in a study with 40 participants~\cite{mark2016neurotics}.
Large-scale interaction data have been used to gain insights into human behavior in the areas of mood rhythms~\cite{golder2011diurnal}, diet~\cite{west2013cookies}, 
conversation strategies~\cite{althoff2016counseling}, 
social networks and mobile games encouraging health behaviors~\cite{althoff2017onlineactions,althoff2016pokemon,shameli2017competition}, 
and health and disease-related search behaviors~\cite{paparrizos2016screening,white2016early}.

\xhdr{This Work}
Existing research on sleep and performance is either small-scale and laboratory-based~\cite{lim2010meta} or relies on subjective measures such as surveys capturing ``typical'' sleep~\cite{sternberg2013largest} which do not allow for temporal coordination of sleep and performance measurements.
As a complement and extension of research to date on performance in artificial laboratory settings, we study real-world cognitive performance which we measure through interactions with a web search engine. 
We use objective measurements of sleep (time in bed) from wearable devices which are preferred to subjective self-reports that can be significantly biased~\cite{lauderdale2008self} and that enable us to study performance variation over time in reference to sleep timing. 
This work represents the largest study of objectively measured sleep and real-world performance to date, employing a subject pool that is orders of magnitude larger than the largest comparable prior  study~\cite{lim2010meta}.
Our study demonstrates on a large scale that interactions with devices are influenced by biological processes and sleep.

\begin{table}[t]
\centering
\resizebox{.85\columnwidth}{!}{%
\begin{tabular}{lr}
  \toprule
   \textbf{Dataset Statistics} &  \\
   \midrule
   Observation period & 18 months\\
    \# users & 31,793\\
    \# nights of sleep tracked & 3,102,209\\
    \# queries & 24,590,345\\
    \# filtered queries with clicks & 6,906,791\\
    \# keystrokes extracted & 68,779,113\\
    \# total interactions & 75,685,904\\
    Average keystroke time & 225ms\\
    Average click time & 9.28s\\
    Median age & 38\\
    \% female & 6.1\% \\
    \% underweight (BMI < 18.5) & 1.4\%\\
    \% normal weight (18.5 $\leq$ BMI < 25) & 32.4\% \\
    \% overweight (25 $\leq$ BMI < 30) & 39.2\%\\
    \% obese (30 $\leq$ BMI) & 27.0\%\\
    Median time in bed & 7.26h\\
  \bottomrule
 \end{tabular}
 }

 \caption{
    Dataset statistics.
    BMI refers to body mass index.
 }
 \label{tab:dataset_statistics}
 \end{table}

\begin{figure}[t]
\centering
\includegraphics[width=1.0\columnwidth]{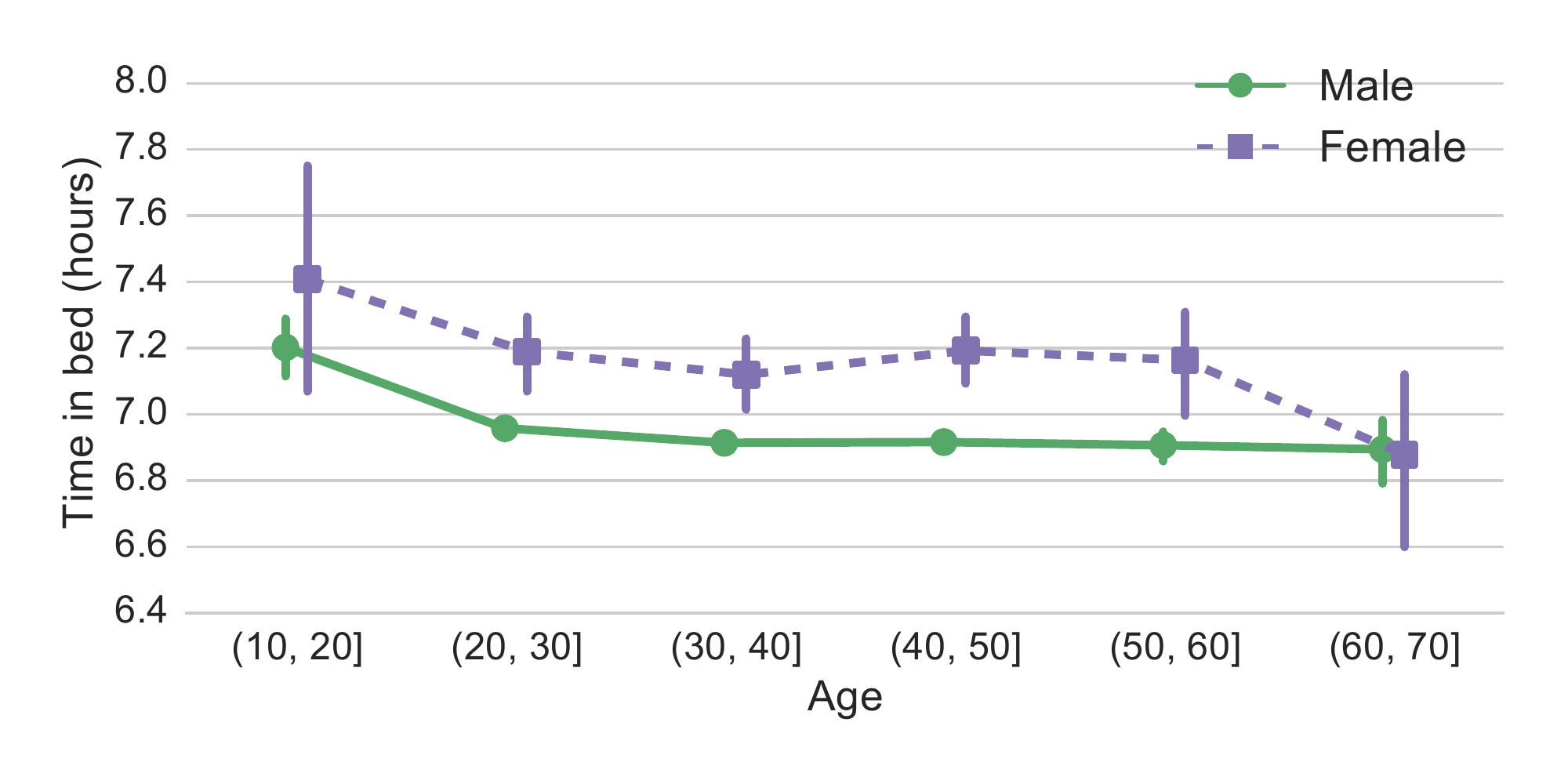}
\caption{
Average sleep duration across age and gender.
Our measurements are consistent with previous estimates~\cite{basner2007american,timeusesurvey2015sleepduration,Walch2016} (Section~\ref{sec:dataset}).
Error bars in all figures correspond to 95\% confidence intervals of the corresponding mean estimates.
}
\label{fig:sleep_duration_by_age_gender}
\end{figure}

\section{Dataset}
\label{sec:dataset}

Our dataset contains over 75 million search engine interactions and sleep measurements for 31,793 US users of Microsoft products 
who agreed to link their Bing searches and Microsoft Band data for use in generating additional insights or recommendations about their sleep or activity. 
Basic dataset statistics and demographic information on the users are summarized in Table~\ref{tab:dataset_statistics}.
Demographic variables (age, gender, body mass index) are self-reported through the Microsoft Health app.
While the user age and overweight/obesity status closely track official estimates in the United States, we note that our sample is predominantly male.

\xhdr{Performance}
We measure performance through the timing of two types of interactions with a search engine (Microsoft Bing): (1) individual keystrokes within the search box that are tracked by the search engine so it can automatically suggest query completions, and (2) clicks on the result page after a search query. 
Section~\ref{subsec:performance_measure_definition} provides more details on each of these measures and we discuss how to account for potential confounds such as the type of query in Section~\ref{subsec:model_estimates_conceptual_model}. 
We exclude search engine interactions originating from mobile devices since such interaction patterns and timing are fundamentally different from those on desktop devices. 
While users could potentially access the search engine from multiple machines, we note that for most users this is unlikely to be the case and that using different keyboards and mice throughout the day is unlikely to explain the timing differences observed in this work. 

\xhdr{Sleep}
Sleep data from wearable devices provides objective measurements which have been preferred to subjective self-reports that may be significantly biased~\cite{lauderdale2008self}. To estimate sleep, we consider signals from wrist-worn activity trackers (Microsoft Band) that include a 3-axis accelerometer, gyrometer, and optical heart rate sensor.
The Microsoft Band employs internally validated proprietary algorithms for estimation of sleep and we focus on duration of time in bed (herein referred to as ``sleep duration''). 
Time in bed is delineated either by manual input of the user (\ie, explicit taps on the device before going to sleep and immediately after waking up) or automatically based on movement if the user does not provide manual input.
The use of an event marker to denote bed timing is widely used in sleep research in lieu of or in concert with sleep diaries~\cite{ancoli2003role}.
Following standard practice~\cite{Walch2016}, we exclude any sleep duration measurements below 4 and above 12 hours of time in bed.

As evidence that our sleep measurements have face validity, we show that they match published sleep estimates.
Figure~\ref{fig:sleep_duration_by_age_gender} illustrates average time in bed across age and gender.
Time in bed decreases with age and is higher in females than males consistent with published estimates~\cite{basner2007american,timeusesurvey2015sleepduration,Walch2016}. 
Walch et al.~\cite{Walch2016} report very similar times and a difference of 17 minutes between females and males. 
With the exception of 60 to 70 year old subjects, we find differences between 12 and 17 minutes.
There is no difference for older subjects, which matches survey-based estimates by Basner et al.~\cite{basner2007american}.
We take these alignments with published research as evidence for the validity of using wearable device-based sleep data for large-scale population studies of sleep and performance.

\begin{figure*}[t]
\centering
\includegraphics[width=1.0\textwidth]{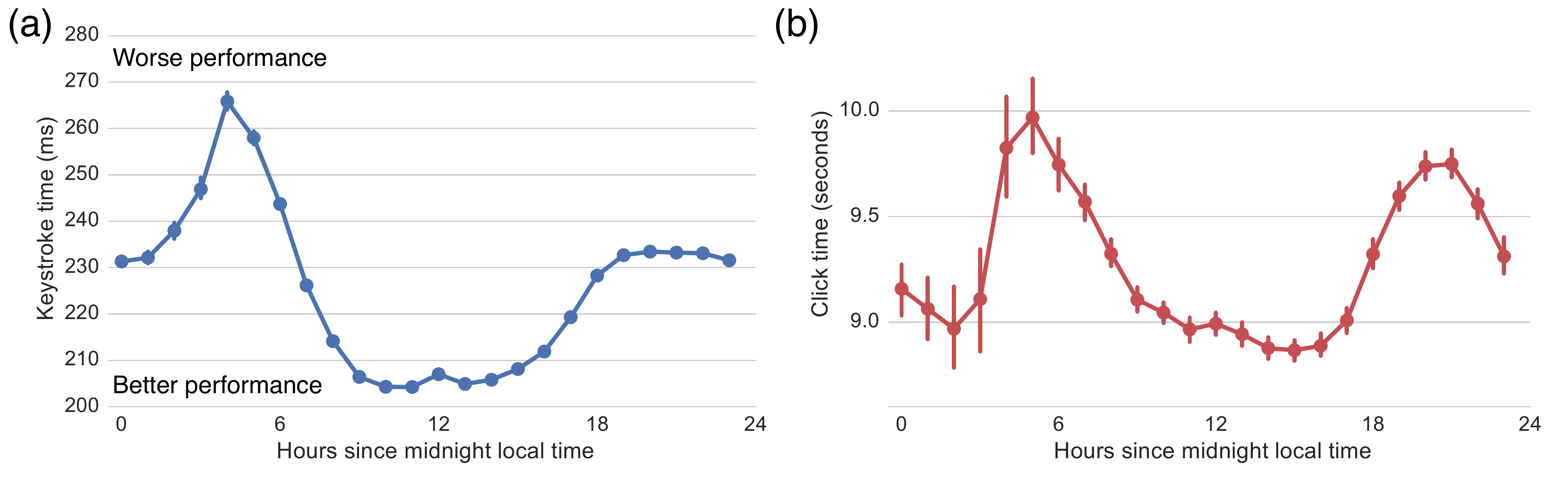}
\caption{
Time of day-dependent variation in keystroke (a) and click timing (b). 
Higher values indicate worse performance.
Both the shape of temporal variation with fastest performance a few hours after wake and slowest performance during habitual sleep times as well as the magnitude of variation are consistent with controlled laboratory-based studies~\protect\cite{ackerman2008sex,dijk1992circadian,dinges1995overview,wise2009handbook} (Section~\ref{subsec:performance_temporal_variation}).
}
\label{fig:raw_timings_by_local_time}
\end{figure*}

\begin{figure}[t]
\centering
\includegraphics[width=1.0\columnwidth]{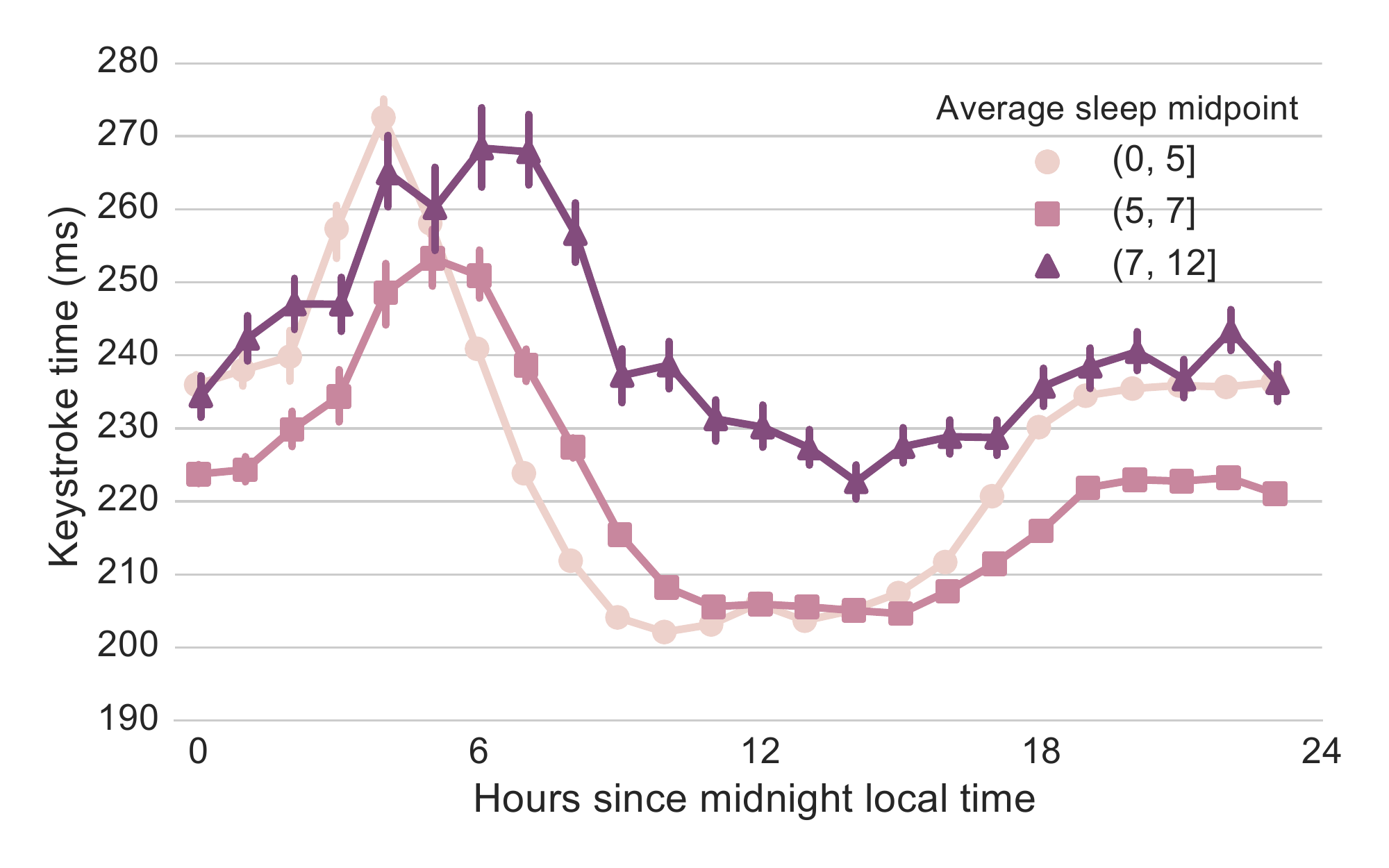}
\caption{
Variation in keystroke time throughout the day varies with chronotype (morning/evening preference) which is defined based on the average point of mid sleep (Section~\ref{subsec:performance_chronotype}).
Users that typically sleep early (light color) perform slowest at about 04:00~h, while medium or late sleepers (darker colors) perform slowest at 05:00~h and 06:00-07:00~h, respectively.
This closely matches their habitual sleep time and is consistent with controlled laboratory-based studies~\protect\cite{matchock2009chronotype}.
}
\label{fig:raw_timings_by_local_time_chronotype}
\end{figure}

\section{Performance Measures Based on Interactions during Search}
\label{sec:search_performance_measures}

Next, we describe two human performance measures derived from search engine interactions that we use to study daily variation in  performance.
We show how these measures exhibit variations in performance over time and based on chronotype (morning/evening preference) consistent with findings from laboratory-based sleep studies.
This demonstrates that performance signals generated from everyday search engine interactions vary based on biological processes. 
We model these processes and influences explicitly in Section~\ref{sec:performance_modeling}.


\subsection{Performance Measures}
\label{subsec:performance_measure_definition}
We study two real-world performance measures in this work since it is possible that different measures would respond differently to sleep deprivation as sleep studies have shown differential effects of sleep deprivation on different measures of cognition. 

\xhdr{Keystroke Time}
The first measure is based on keystroke timing. 
The search engine's search box registers every single keystroke and sends a request for query completions to the search engine's servers.
We use the timing between two such requests as the time of a single keystroke if the two queries are different by exactly one character (not every request is received on the server side) and within two seconds (larger times indicate longer thought processes or separate sessions).
This threshold is sensible as an average keystroke by an average typist takes about 240 milliseconds (50 words per minute at 5 characters per word~\cite{card1980keystroke}).

\xhdr{Click Time}
The second measure is based on the time to click on a search result after a search result page is displayed.
We measure the time between the search query and the first click on any result on the first page.
Click times over two minutes are excluded since they might stem from interrupted sessions. 
We account for click position and query type as described in Section~\ref{subsec:model_estimates_conceptual_model}.

We believe that investigating measures that capture performance on two different tasks provides robustness and breadth to our analyses. 
The two tasks rely on different mixes of sensing, reflection, planning, and formulating, executing, and monitoring of motor plans~\cite{pilcher1996effects}. 
Studies of the potential subprocesses for each task and how they might be differentially influenced by sleep is beyond the scope of this paper.  
However, our search engine interactions capture performance in everyday tasks that are highly relevant to many occupations, as captured by typing and searching for information~\cite{Pew2011searchusers},
and allow us to non-intrusively measure changes in real-world performance throughout the day.

Note that all timing measurements are taken on the server side and not the client side. 
Therefore, it is important to consider the potential influence of network latency factors.
We found that the network latency changes only very little between two consecutive requests (less than 1 millisecond) and thus any latency effects cancel out when we take the time difference between two requests (details in online appendix~\cite{althoff2016onlineappendix}).
This demonstrates that variation in network latency does not affect our analyses. 
Furthermore, variations in site rendering time (\ie, measuring time from first script till page load completed including dynamic contents) are much smaller (order of milliseconds) compared to variation in click times.


The temporal variation sensed in performance could potentially be an artifact of different users contributing timings at different time points instead of actual within user variation throughout the day.
However, we verified that the temporal variation we observe is due to within user variation throughout the day by confirming that the patterns of temporal variation are effectively identical for raw measurements and within-user normalized variants (Z-scores; online appendix~\cite{althoff2016onlineappendix}). We also verified that performance variation during the weekend is similar to variation during the week (online appendix~\cite{althoff2016onlineappendix}) and we therefore do not further differentiate between performance during weekdays and weekends in this paper.
Finally, we considered alternative performance measures based on backspace usage in keystrokes and spelling errors in search queries.
Since we found results to be similar to keystroke and click timing but more noisy due to less frequent measurements, we report results on keystroke and click timing in this paper.

\subsection{Temporal Variation of Keystroke and\\Click Times}
\label{subsec:performance_temporal_variation}
Next, we validate our methodology by considering the findings obtained from small-scale controlled sleep studies.
It is well established that human performance varies over time and follows a circadian rhythm~\cite{ackerman2008sex,wise2009handbook}.
Keystroke and click timing also vary throughout the day in a daily rhythm as illustrated in Figure~\ref{fig:raw_timings_by_local_time}.
Keystroke times (Figure~\ref{fig:raw_timings_by_local_time}a) are on the order of 240 milliseconds which closely matches the expected typing speed of an average typist (240 milliseconds; 50 words per minute at 5 characters per word, see~\cite{card1980keystroke}).
Click times (Figure~\ref{fig:raw_timings_by_local_time}b) are on the order of 10 seconds.
Note that both measures follow a similar pattern throughout the day.
Users are fastest to type and click a few hours after typical wake times and the timing increases again in the evening hours (in particular for click times).
Performance is slowest during habitual sleep times (\eg, 04:00~h) closely matching accident risk rates~\cite{dinges1995overview}
and the anticipated circadian nadir (\ie, the time of greatest circadian sleep drive)~\cite{dijk1992circadian}.  
Furthermore, controlled laboratory experiments have shown that performance typically varies by 15 to 30 percent over the course of a day across a variety of simple motor and cognitive tasks~\cite{ackerman2008sex,wise2009handbook}.
For keystrokes we measure a variation of 31\% and for click times a variation of 12\%.

The consistent agreement in shape and magnitude of variation with controlled lab experiments on human performance and for two different tasks suggest that these large-scale measures based on search engine interactions can be used to study sleep and performance.
The proposed measures can be collected non-intrusively at unprecedented scale and shine light on how real-world performance varies throughout the day and with changes in sleep.

\subsection{Performance Variation by Chronotype}
\label{subsec:performance_chronotype}

A person's chronotype encompasses the propensity for the individual to sleep at a particular time during a 24-hour period and is at least partly based on genetics~\cite{roenneberg2003life}. 
Studies have shown that performance depends on the alignment of chronotype and time of day~\cite{matchock2009chronotype}; early types tend to be higher performing earlier in the day while late types are higher performing later.
The individual chronotype of each user can be defined based on the mid-sleep point on free days ($MSF$) which is the halfway point between going to sleep and waking up~\cite{juda2013chronotype,roenneberg2003life}.
Many people compensate for slept debt accumulated during work days by sleeping longer on free days; that is, the sleep midpoint we observe is later than the internal biological clock would dictate on the free days.
Therefore, sleep scientists use a midsleep point that is corrected for oversleep (indicated by $SC$)~\cite{juda2013chronotype}: 
$MSF_{SC} = MSF - 0.5 (SD_F - (5*SD_W + 2*SD_F)/7)$, where $SD_F$ and $SD_W$ are sleep duration and free days and work days, respectively, and $SD_F - (5*SD_W + 2*SD_F)/7$ corresponds to the difference in sleep duration on free days and the average day.
We compute this corrected midpoint for every user in the dataset using weekdays as work days and weekend days as free days (Median $MSF_{SC}=4.70$).

We show that keystroke times throughout the day vary with chro\-no\-type (Figure~\ref{fig:raw_timings_by_local_time_chronotype}), matching results from previous sleep studies~\cite{matchock2009chronotype} and thus providing further validation of our methods.
We find that early sleepers are slowest at about 04:00~h, while medium or late sleepers are slowest at 05:00~h and 06:00-07:00~h, respectively. 
This closely matches each group's habitual sleep time and demonstrates the validity and power of this large dataset; for each chronotype group, we have millions of measurements even during typical sleep times that allow us to estimate these performance curves.
We find similar results for click times.

\section{Modeling Performance}
\label{sec:performance_modeling}

Having demonstrated that performance of search engine interactions vary over time and based on biological processes (Section~\ref{sec:search_performance_measures}), we now operationalize and extend a conceptual model of sleep and performance from chronobiology~\cite{aakerstedt1997three,borbely1982two} to explain the variation observed in performance measurements. 
Classic sleep models are based on circadian rhythms and homeostatic sleep drive~\cite{borbely1982two}.
In addition, we consider sleep inertia and sleep duration~\cite{aakerstedt1997three,van2000circadian}.
Background on relevant biological processes is covered in Section~\ref{sec:related_work}.

\begin{figure*}[h!]
\centering
\includegraphics[width=1.0\textwidth]{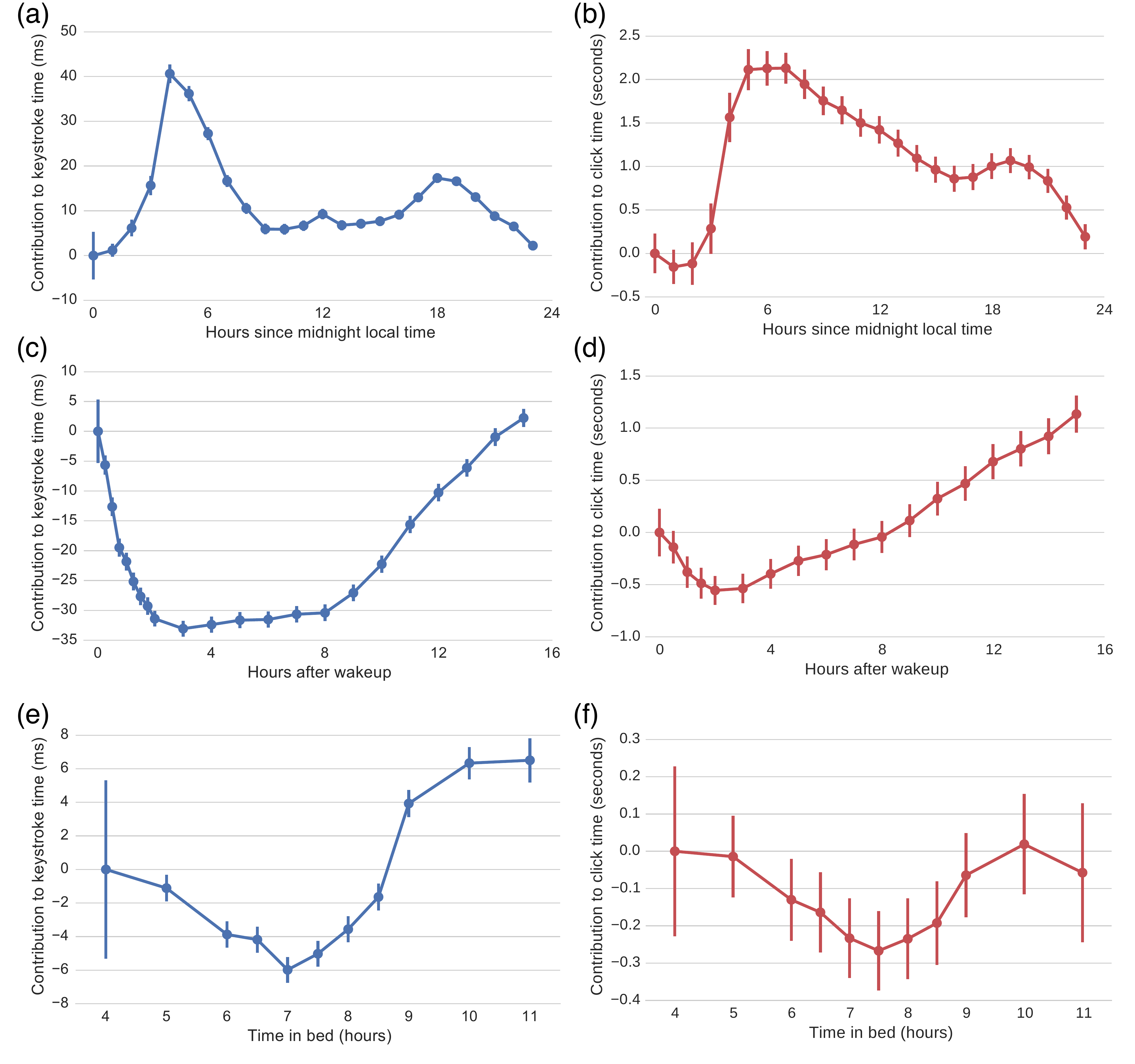}
\caption{
Contributions to keystroke (a,c,e; blue) and click time (b,d,f; red) performance of different factors included in our model.
Results are similar for both performance measures and match estimates from controlled sleep studies in the laboratory (Section~\ref{sec:performance_modeling}).
For example, variation over the time of day $\mathbf{c^t}$ (a,b) shows that performance is slowest during habitual sleep times near the presumptive circadian nadir (04:00~h; see main text).
Variation across time after wake up $\mathbf{c^w}$ (c,d) shows effects of sleep inertia during the first two hours after wake. There is relative stability for around eight hours in keystroke time but a steady decline in click time after that point.
Sleep durations $\mathbf{c^d}$  (e,f) of 7.0-7.5 hours are associated with optimal performance according to our measures.
However, note that the impact on overall variation is smaller compared to time of day (a,b) and time since wake up (c,d). 
}
\label{fig:model_estimates}
\end{figure*}

\subsection{Conceptual Model}
\label{subsec:model_estimates_conceptual_model}
We model the keystroke and click timing based on (1) time of day in local time,
(2) time in hours after wake up, and (3) sleep duration the previous night. 
We know (1) from the time of the keystroke or click time measurement, and (2) and (3) from wearable device-defined sleep measurements (Section~\ref{sec:dataset}).

Since many people wake up during the same morning hours every day, time of day and time since wake up are naturally correlated and challenging to disentangle. 
In laboratory-based sleep studies, the goal of exploring the distinct influences of the factors is achieved by ``forced desynchrony'' protocols~\cite{van2000circadian}, where subjects are deprived of sleep for extended periods of time. 
Instead of similar interventions, we employ mathematical modeling with a large-scale dataset of real-world sleep and performance measurements and use the variation observed across millions of observations to disentangle the relative contributions of circadian and homeostatic factors.
The large-scale dataset contains numerous performance measurements during usual (day) and unusual (late night) times (\eg, Figure~\ref{fig:raw_timings_by_local_time_chronotype}) that we can use to understand the relative contributions of these factors to performance in the open world (see formulation of additive model in Section~\ref{subsec:model_estimates_mathematical_formulation}).

\xhdr{Potential Confounding Factors}
We control for several factors in our model to avoid confounding. For keystrokes, we control for the exact character typed or removed since different characters might take a varying amount of time (\eg, typing an ``a'', or a capital ``A'', or hitting backspace).
For click times, it is expected that clicking on results further down the list of results will take more time, which holds true in our data (online appendix~\cite{althoff2016onlineappendix}). 
We therefore control for the click position in our model.

Clicking on a result link is preceded by a cognitive process--interpreting the words displayed on links and deciding which link to click--which can be quick in the case of navigational queries (\eg, ``facebook'') or much slower in the case of informational queries (\eg, ``What is the homeostatic sleep drive?'').
Formally, this distinction can be captured through the concept of click entropy, which measures how ``surprising'' the distribution over clicked URLs for a given query is~\cite{dou2007large}.
We find that informational queries take about two seconds longer than navigational queries on average (online appendix~\cite{althoff2016onlineappendix}).
Therefore, we control for the click entropy of the query preceding the click in our model.

An extreme way of controlling for varying queries is to compare click times for exactly identical queries (\eg, popular queries such as ``facebook'').
We verified that this yields very similar results, 
albeit with larger confidence intervals since the sample size is reduced dramatically compared to including all queries and controlling for click entropy, 
demonstrating that the observed patterns are not due to a particular mix of query types.

In addition, we tested for learning effects as issuing the same query multiple times might lead to improved performance.
However, most queries, 73.1\%, are unique in the dataset and only 4.1\% of queries occur more than three times.
Further, we did not find any evidence for improving performance over time for frequently occurring queries.
This is likely because most users were fairly proficient at typing before the start of our observation period.

\subsection{Mathematical Formulation}
\label{subsec:model_estimates_mathematical_formulation}
We now describe the formulation of the model for keystroke timing. The model for click times is parallel, where we control for the click position and click entropy instead of the keystroke type.
We are interested in estimating how (1) time of day, (2) time after wake up, and (3) sleep duration influence performance.
We assume that all these effects are additive as supported by evidence presented in~\cite{achermann1994simulation}.
Mathematically, we formulate a fixed-effects model 
$$y_i = \alpha + f^k(x_i^k) + f^t(x_i^t) + f^w(x_i^w) + f^d(x_i^d) + \epsilon_i \, ,$$
where $y_i$ is the keystroke time for observation $i$, $\alpha$ is a constant intercept, and $f^k, f^t, f^w, f^d$ are the unknown functions of interest for keystroke type, time of day, time since wake up, and sleep duration, respectively, with corresponding input features $x_i^k, x_i^t, x_i^w, x_i^d$, and $\epsilon_i$ is the $i$-th residual.

Instead of estimating arbitrary functions, we use fine-grained piecewise constant approximations.
We discretize each input space (\eg, between midnight and 01:00~h, or between 01:00~h and 02:00~h, or between 0 and 15 minutes after waking up, \etc).
We denote the functions mapping input features $x_i^t, x_i^w, x_i^d$ to their respective bins as $b^t, b^w, b^d$ (note that keystroke type $x_i^k$ is already discrete).
Further, we use the functions $c^k, c^t, c^w, c^d$ to map the discretized features to a constant value.
The simplified model then becomes
$$y_i = \alpha + c^k(x_i^k) + c^t(b^t(x_i^t)) + c^w(b^w(x_i^w)) + c^d(b^d(x_i^d)) + \epsilon_i  \, .$$
The outcome of interest in this modeling task are the functions $c^t, c^w, c^d$ which express the independent impact of (1) time of day, (2) time since wake up, and (3) sleep duration on performance timings the next day.
We estimate all parameters $(\alpha, c^k, c^t, c^w, c^d)$ including 95\% confidence intervals through least squares optimization.
We also experimented with mixed effects models controlling for variation across users and across queries through random effects.
While standard mixed model libraries do not scale well to the size of our dataset, we found that these models lead to very similar estimates compared to the fixed effects model described above when using subsets of the data.

\subsection{Results}
\label{subsec:model_estimates_results}
The functions $c^t, c^w, c^d$ modeling the influence on cognitive performance of time of day, time since wake up, and sleep duration are illustrated in Figure~\ref{fig:model_estimates}. 
Impact on keystroke timings are shown in blue (Figure~\ref{fig:model_estimates}a,c,e) and impact on click times are shown in red (Figure~\ref{fig:model_estimates}b,d,f).
Note that the shapes of these functions for keystrokes and click times are very similar and smooth, even though there are no constraints that would force this to occur.
Furthermore, we note that the temporal variation in cognitive performance is not explained by variation in different users that contribute timings at different points throughout the day (\ie, population differences) but are due to within user variation (online appendix~\cite{althoff2016onlineappendix}).

\xhdr{Time of Day}
Cognitive performance on both keystroke and click tasks varies with time of day (Figure~\ref{fig:model_estimates}a,b) and is slowest during habitual sleep time around 04:00-06:00~h. 
Performance quickly improves after typical wake times and becomes slightly slower in the evening for both keystroke and click times (19:00~h).
The two curves consistently match estimates of circadian rhythm processes in sleep obtained through controlled laboratory experiments~\cite{dijk1992circadian,wright2012circadian}.
Note that the magnitude of variation is substantial at around 40 milliseconds for keystrokes and over 2.1 seconds for click times, which are changes of 18\% and 23\%, respectively, relative to average timing for each (Table \ref{tab:dataset_statistics}).

\xhdr{Time after Awakening}
Cognitive performance also varies substantially with the time after wake up (Figure~\ref{fig:model_estimates}c,d). 
The magnitude of the variation is relatively large at about 42 milliseconds or 19\% for keystrokes about slightly over 1.6 seconds or 17\% for click times.
Within the first two hours, performance rapidly improves (\ie, lower timings).
This demonstrates a well-known effect in sleep studies called sleep inertia (Section~\ref{sec:related_work}).
After this point, performance is best and slowly worsens until a point of poorest performance is reached at around 16 hours of wake time, consistent with the homeostatic sleep drive~\cite{borbely1982two}. 
This corresponds exactly to the point when most people would go to sleep again (\ie, a typical sleep duration of 8 hours).
We excluded data beyond the typical wake period of 16 hours because the data becomes more sparse and to avoid potential selection effects with regard to the people who choose to stay awake for exceptionally long periods of time.  
However we found similar patterns between both keystrokes and click times even beyond this point. 
We note that keystroke time is relatively stable for about six hours while click times continuously increase, likely due to the differences in cognitive and motor competencies for the tasks, and due to differences in the sensitivities of those competencies to status of sleep and circadian rhythm.
In summary, the estimates derived from our model closely capture the initial sleep inertia and the increasing homeostatic sleep drive first discovered through laboratory-based studies~\cite{aakerstedt1997three,van2000circadian,wright2012circadian}.

\xhdr{Time in Bed}
Keystrokes and click time vary with the amount of time in bed during the previous night (Figure~\ref{fig:model_estimates}e,f).
However, we note that this variation, 12 milliseconds for keystrokes (5\%) and 0.25 seconds for click times (3\%), is much smaller than the previous two factors.
For both measures, we find a clear U-shaped curve with its center, indicating optimal performance, at 7.0-7.5 hours of sleep.
Both sleeping too little (under 7 hours) or too much (more than 8-9 hours) are associated with decreased performance.
U-shaped relationships with respect to sleep duration have been reported for several outcomes (\eg, mortality~\cite{kripke1979short}).
We further investigate the impact of insufficient sleep on performance in Section~\ref{sec:bad_sleep_impact}.

\begin{figure*}[th!]
\centering
\includegraphics[width=1.0\textwidth]{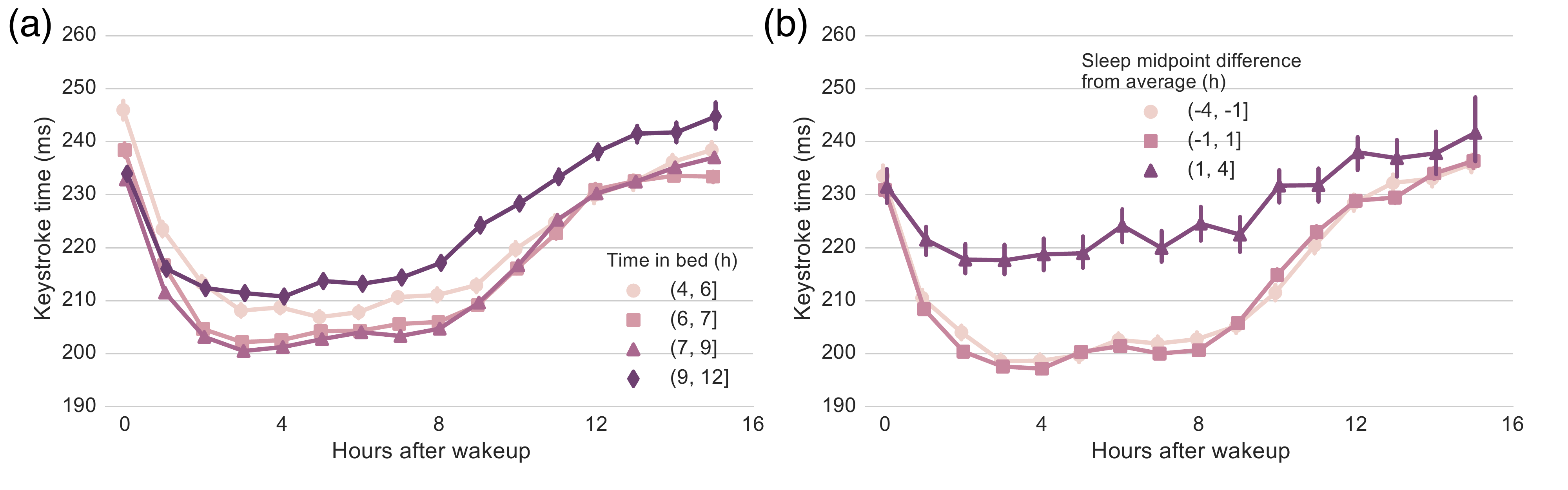}
\caption{
The impact of sleep duration (a) and timing (b) on performance the next day.
Sleep timing is measured through difference from the typical sleep midpoint and we control for sleep duration.
We find that sleeping less than 7 or more than 9 hours is associated with slower performance (a). 
Sleeping earlier than usual does not make a large difference but going to bed an hour or more later than usual is associated with significantly worse performance the next day (b).
}
\label{fig:bad_sleep_single_night}
\end{figure*}

\begin{figure}[!th]
\centering
\includegraphics[width=1.0\columnwidth]{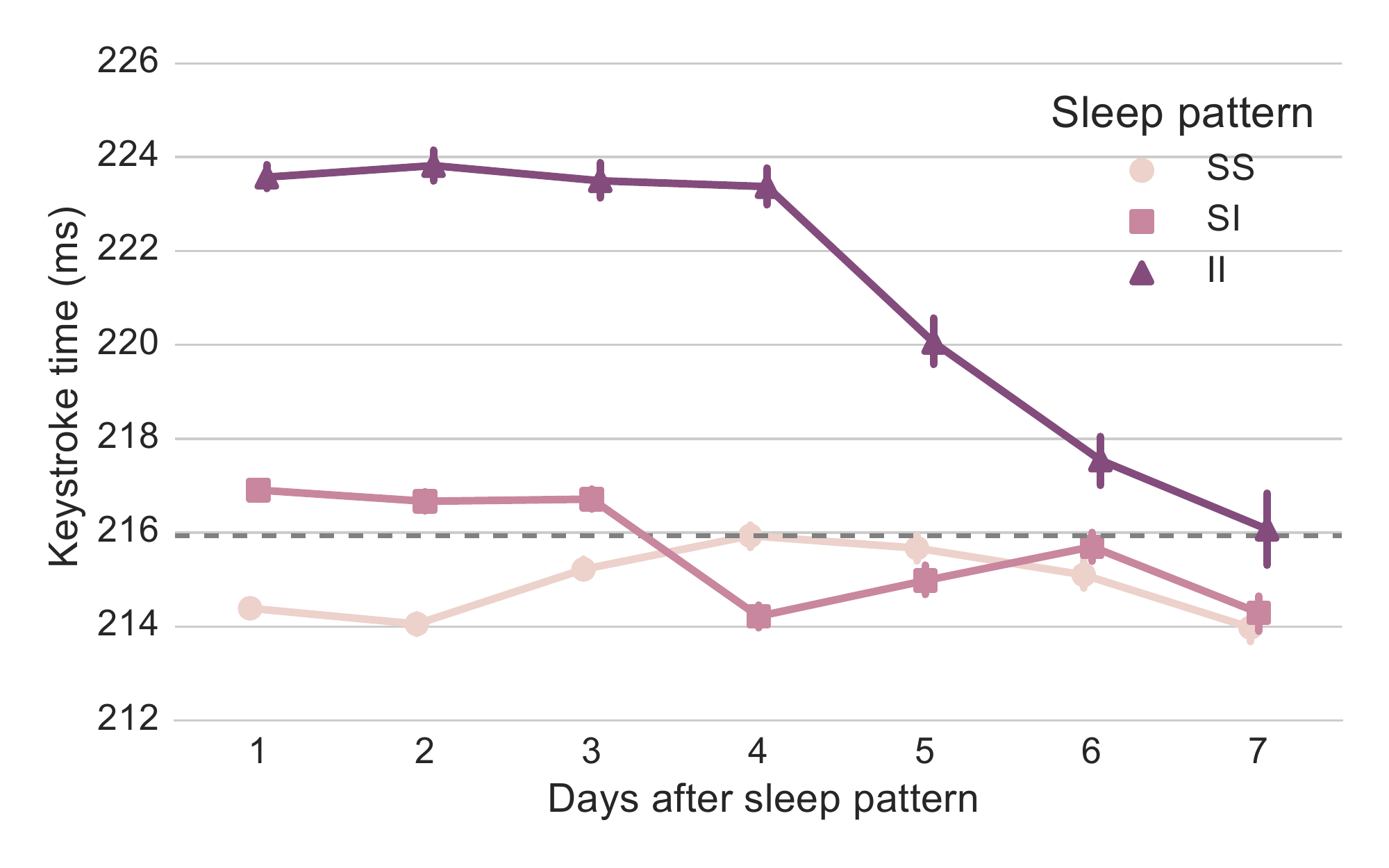}
\caption{
Comparing the impact on performance of zero (SS), one (SI), or two (II) consecutive insufficient nights of sleep (less than six hours of time in bed). 
One night of insufficient sleep is associated with significantly slower keystroke times and 
two insufficient nights in a row exhibit a significantly larger effect.
Judging by when average keystroke time drops below the horizontal dashed line representing the slowest performance for the group with two nights of sufficient sleep (SS), 
we observe that it takes six nights of sleep to return to baseline performance levels after two nights of insufficient sleep (day 7) and three nights to return to baseline performance levels after one night of insufficient sleep (day 4) given real-world sleep schedules.
}
\label{fig:bad_sleep_two_nights}
\end{figure}

\section{Influence of Insufficient Sleep on Performance}
\label{sec:bad_sleep_impact}

Following our studies to validate the methodology (Section~\ref{sec:search_performance_measures} and Section~\ref{sec:performance_modeling}), we now extend prior laboratory-based sleep research with estimates of how sleep influences performance in real-world settings. 
In particular, we study the impact of one or multiple nights of insufficient sleep on performance over the following days.

\subsection{Single Nights of Insufficient Sleep}
We first consider single nights of sleep and analyze how very short or very long sleep durations, as well as differences in sleep timing from the usual patterns within a user, impact performance. 
We only show results for keystroke timing here; the results are similar for click times (\eg, Figure~\protect\ref{fig:raw_timings_by_local_time} and Figure~\protect\ref{fig:model_estimates}).
Figure~\ref{fig:bad_sleep_single_night}a shows that users performed significantly slower when in bed fewer than 6 or more than 9 hours, consistent with the results described in Section~\ref{subsec:model_estimates_results}.
In those conditions, the average keystroke times were about four and seven milliseconds longer compared to sleeping between 7 and 9 hours (increases of 2.7\% and 4.0\%, respectively; both $p \ll 10^{-10}$; Mann--Whitney U-test, which is used for all hypothesis tests in this section). 

Timing of sleep is also a significant factor for performance the next day (Figure~\ref{fig:bad_sleep_single_night}b).
While sleeping earlier than usual makes only a difference of about 1 milliseconds or 0.5\% ($p \ll 10^{-10}$), going to bed an hour or more later than usual is associated with significantly worse average performance of about 14 milliseconds or 7.3\% longer keystrokes ($p \ll 10^{-10}$). 
Note that we limited the sleep duration to be between 7 and 8 hours long for this analysis so that these results demonstrate the impact of timing independent of differences in duration (\ie, those going to sleep later had a normal length of time in bed despite going to sleep late).
We further verified that these results are not due to people sleeping later and longer on weekends when they might be typing slower due to less work pressure as we find similar patterns and effect sizes using just weekday data.
Thus, these results could point to an interaction between the circadian clock and the ultradian rhythm of sleep (\ie, the cycling of sleep stages): sleeping at different phases can result in different sleep organization~\cite{dijk1995contribution}.
Our findings suggest that sleeping later in one's circadian cycle does not satisfy the neural recovery needed for proper daytime performance, while sleeping earlier does not have the same negative effects.

\subsection{Multiple Nights of Insufficient Sleep}
Above, we reported on the effect of a \textit{single} night of sleep with particular duration and timing on the next day.
Here, we examine whether \textit{multiple} insufficient nights of sleep measurably affect performance and how long this effect appears to persist.
For purposes of this analysis, we define an ``insufficient'' night of sleep (``I'') to have a time in bed of under six hours (as in~\cite{fernandez2010insomnia}), and a ``sufficient'' night of sleep (``S'') to have a time in bed of at least six hours. 
We consider three different scenarios:
two nights of sleep with more than six hours each (SS), one night over and the next night under six hours (SI), and two nights under six hours of sleep (II). 
We measure the performance after those two nights of sleep for a period of seven days, reducing the performance on each of these seven days to a single value---the average performance during the first 16 hours after wake up (\ie, typical wake period).
We do not consider longer sleep patterns here due to the large number of possible combinations and data reduction associated with individual sleep patterns (\eg, a person might not track their sleep every single night).
Intentionally not controlling for sleep both preceding and following the two nights of interest, we are addressing how insufficient sleep impacts real-world performance given real-world choices. 
We are not, however, examining the underlying biological processes of recovery from sleep loss.
We note that the start of the sleep patterns was distributed all throughout the week; for example, two nights of sufficient sleep (SS) did occur both during the week as well as over the weekend. 
We define recovery time as the number of days it takes to reach performance levels comparable to those after a sufficient sleep schedule (SS).

\xhdr{Results}
Multiple insufficient nights of sleep have a significant impact on average keystroke timing (Figure~\ref{fig:bad_sleep_two_nights}).
Performance is best after two sufficient nights of sleep, slightly but measurably worse after one insufficient night of sleep, and significantly worse after two insufficient nights in a row.
Over the first 24 hours, having one insufficient night of sleep is associated with 1.2\% slower performance ($p \ll 10^{-10}$) and two insufficient nights of sleep are 4.8\% slower ($p \ll 10^{-10}$) compared to two nights with longer than six hours of sleep each (2.7\% and 7.3\% increases for click times, respectively; both $p \ll 10^{-10}$). 
Note that these effect estimates take into account any real-world behavioral compensation such as increased caffeine intake that will help improve performance after sleep loss. 
The horizontal dashed line in Figure~\ref{fig:bad_sleep_two_nights} corresponds to the slowest keystroke time after two nights of sufficient sleep (SS), 
which we use as a conservative point of reference to judge when performance after insufficient sleep (SI and II) has returned to a performance below this point.
We find that, on average, it takes three nights to make up one insufficient night of sleep (SI crosses dashed line on day 4) and six nights two make up two insufficient nights of sleep in a row (II crosses dashed line on day 7). 
We find very similar results for the impact on the \textit{variance} (\ie, instead of mean) of keystroke timing as well as for click times. 
A version of Figure~\ref{fig:bad_sleep_two_nights} that visualizes average performance throughout each of the seven days is included in the online appendix~\cite{althoff2016onlineappendix}.

Note that these results are not simply due to having fundamentally different users contribute to each of the the curves (SS, SI, II). 
While some users are more likely to get fewer than six hours of sleep than others, we do find similar effects by restricting each of the three curves to be estimated from the exact same set of users.
We note that, since we enforce no constraints on time in bed during the seven days following the sleep pattern, additional nights of insufficient sleep could occur during the follow-up period, contributing to the duration of the recovery period. Thus, we need to explore whether there is a higher likelihood of sleep deficiencies on days following the initial observed two-day period of insufficient sleep. 
We find that, on average, SS is followed by 0.4 nights of insufficient sleep during the following seven days, whereas SI and II are followed by 1.2 and 2.5 such nights. 
Thus, additional days of insufficient sleep for the SI and II cases may have an influence on the overall time to returning to baseline performance. 
Nevertheless, our findings show real-world timing of return to baseline performance.  We leave to future work the study of more complex real-world patterns of sleep and sleep deficit and the influences of sleep deficits on performance.

\section{Conclusion}
\label{sec:conclusion}


Understanding human performance and its relation to sleep is critical to productivity~\cite{colten2006sleep}, learning~\cite{kelley2015synchronizing}, and avoiding accidents \cite{colten2006sleep,dinges1995overview}.
Human performance is not constant but exhibits daily variations~\cite{van2000circadian}.
Existing research on sleep and performance has typically been restricted to small-scale laboratory-based studies involving artificial performance tasks in an artificial environment.
Therefore, novel methods of large-scale real-world monitoring, like we have presented, are necessary to advance our understanding of sleep and performance~\cite{roenneberg2013chronobiology}.

\xhdr{Summary of Results}
We presented the largest study to date on sleep and performance in the wild.  Using a new approach to non-intrusive measurement for both cognitive performance and sleep we were able to study more than 400 times the number of users compared to the second largest study.
We correlated human performance based on interactions with a web search engine to sleep measures detected by a wearable device.
We demonstrated that real-world performance varies throughout the day and based on chronotype and prior sleep, in close agreement with small-scale laboratory-based studies.
We developed a statistical model that operationalizes recent chronobiological research and showed that our estimates of circadian rhythms, homeostatic sleep drive, and sleep inertia closely match published results of controlled sleep studies.
Further, we contribute to existing sleep research through quantifying extended periods of lower real-world performance that are associated with single and multiple nights of insufficient sleep.

\xhdr{Implications}
We have demonstrated that human performance can be measured in a real-world setting without any additional hardware or explicit testing by exploiting existing search engine interactions that occur billions of times per day. 
We have validated our methodology and shown that human performance, as measured through these signals, varies throughout the day and based on chronotype and sleep, in close agreement with controlled laboratory-based studies.
Beyond the relevance of the results to extending insights about sleep and performance, our findings more generally highlight the potential power of harnessing online activities to study human cognition, motor skills, and public health. Large-scale physiological  sensing from online data enables
\begin{itemize}
    \item studies of sleep and performance outside of small laboratory settings, and without actively inducing sleep deprivation, 
    \item non-intrusive measurement of cognitive performance without forcing individuals to interrupt their work to perform separate artificial tasks~\cite{roenneberg2013chronobiology},

    \item the identification of realistic measures of real-world cognitive performance based on frequent tasks and interactions, 
    \item and continuous monitoring of such measures.
\end{itemize}
Suitable examples for such data include continuous usage patterns from computing applications such as email, programming environments, bug report systems, office suites, and others.
Any insights on performance and productivity gained through monitoring these applications could be used to improve the user's awareness of such patterns and to 
adapt the user experience appropriately (\eg, scheduling tasks intelligently in order to prevent or minimize human error; scheduling meetings based on participants performance and chronotype profiles).
There are great opportunities ahead to investigate how such insights could be used to personalize applications based on relevant biological processes and chronotypes.


\xhdr{Acknowledgments}
We thank Jure Leskovec, Emma Pierson, Marinka Zitnik, David Hallac,  David Jurgens and the anonymous reviewers for their valuable feedback on the manu\-script.


\balance


\bibliographystyle{abbrv}
\bibliography{refs}



\end{document}